# Calcium-Decorated Graphene-Based Nanostructures for Hydrogen Storage

Hoonkyung Lee,[1,2] Jisoon Ihm,[3] Marvin L. Cohen,[1,2] and Steven G. Louie[1,2,*]

[1]Department of Physics, University of California, Berkeley, California 94720, USA

[2]Materials Science Division, Lawrence Berkeley National Laboratory, Berkeley, California 94720, USA

[3]Department of Physics and Astronomy, Seoul National University, Seoul, 151-747, Korea

*Corresponding author: sglouie@berkeley.edu.

ABSTRACT

We report a first-principles study of hydrogen storage media consisting of calcium atoms and graphene-based nanostructures. We find that Ca atoms prefer to be individually adsorbed on the zigzag edge of graphene with a Ca-Ca distance of 10 Å without clustering of the Ca atoms, and up to six $H_2$ molecules can bind to a Ca atom with a binding energy of ~0.2 eV/$H_2$. A Ca-decorated zigzag graphene nanoribbon (ZGNR) can reach the gravimetric capacity of ~5 wt % hydrogen. We also consider various edge geometries of the graphene for Ca dispersion.

KEYWORDS. Calcium, graphene nanoribbons, and hydrogen storage.

Hydrogen storage in solid-state materials is of importance in many applications, e.g., for the development of hydrogen fuel-cell powered vehicles.[1,2] For the last decade, metal or chemical hydride materials have been considered as a hydrogen storage medium; however, these materials have issues such as slow kinetics and poor reversibility.[1] Recently, carbon-based nanostructured materials have also been explored as a hydrogen storage medium because of the possibility of good reversibility, fast kinetics, and high capacity (large surface area).[3-6] However, it has been found that the storage capacity in these nanomaterials significantly decreases at room temperature and ambient pressure[7,8] because the binding energy of $H_2$ molecules on these materials is short of the desired energy of ~0.2−0.6 eV[9].

In recent years, theoretical studies have suggested two approaches to enhance the interaction of $H_2$ molecules to materials. One approach is to use the induced polarization of $H_2$ molecules by an electric field. Alkali metal or alkali earth metal (Li, Na, and K)-decorated, charged, and Ca-coated nanostructured materials can bind $H_2$ molecules with a binding energy of ~0.1−0.2 eV.[10-15] These materials ideally can reach the gravimetric goal of ~8−9 wt % hydrogen (set by the U.S. Department of Energy (DOE) to be reached by the year 2015[16]). The other approach is to employ the hybridization of $H_2$ σ or σ* orbitals with transition metal $d$ orbitals (the so-called the Kubas interaction[17]). Early transition metal (Sc, Ti, and V)-decorated nanostructured materials adsorb multiple $H_2$ molecules with a binding energy of ~0.2−0.6 eV and may satisfy the DOE goal[18-24]. Experimental studies on hydrogen storage materials employing the Kubas interaction have been reported using reducible mesoporous Ti oxides, organic Ti complexes, and Ti-ethylene complexes.[25-27] However, transition metal atoms basically prefer being clustered to being individually dispersed on nanomaterials because of the large cohesive energy of bulk transition metals (i.e., ~4 eV).[28,29] Clustering changes significantly the adsorption nature of the $H_2$ molecules, causing the $H_2$ molecules to dissociate. So, the hydrogen storage capacity is much lower than what was proposed theoretically for the dispersed geometry.

More recently, the use of calcium atoms has been considered as a decorating element instead of the transition metal atoms because of the low cohesive energy (1.8 eV) of bulk calcium.[30,31] It has been found that clustering of Ca atoms is suppressed on boron-doped or defective carbon nanotubes (CNTs), and the Ca atom adsorbs multiple $H_2$ molecules with a binding energy of ~0.2 eV. Another attractive feature is that the hybridization of the unoccupied Ca $3d$ states with the $H_2$ σ states as well as the polarization of $H_2$ molecules both contribute to the $H_2$ binding to Ca atom.[30] The study shows the feasibility of the hydrogen storage using metal-decorated system.[19-25] However, high concentrations of B dopants or defects are necessary to achieve high Ca coverage on B-doped or defective carbon nanotubes.[30]

In this paper, we show the feasibility of high Ca coverage on graphene-based nanostructures without the aid of any dopants and defects or with a low concentration of B dopants. We find that the Ca atoms prefer to be individually adsorbed on the zigzag edge of graphene and clustering of Ca atoms is suppressed. Up to six $H_2$ molecules can be attached to a Ca atom with a binding energy of ~0.2 eV/$H_2$, and a Ca-decorated ZGNR can attract $H_2$ molecules with the gravimetric capacity of 5 wt % hydrogen. We also consider different edge geometries of the graphene, namely, combined zigzag-armchair-edged graphene nanoribbons (GNRs), large vacancy-defected graphenes with zigzag edge, and boron-doped armchair graphene nanoribbons (AGNRs) for Ca dispersion.

All of our calculations were carried out using a first-principles method based on density functional theory[32] as implemented in the Vienna Ab-initio Simulation Package (VASP) with a projector-augmented-wave (PAW) method[33]. The exchange-correlation energy functional in the generalized gradient approximation (GGA) of Perdew and Wang was used[34], and the kinetic energy cutoff was taken to be 26 ryd. The optimized atomic positions were obtained by relaxation until the Hellmann-Feynman force on each atom was less than 0.01 eV/Å. Supercell[35] calculations were employed throughout where the atoms on adjacent nanostructures are separated by over 10 Å.

To search for hydrogen-storage nanostructured materials consisting of graphene-based nanostructures and calcium atoms, we first consider ZGNRs and AGNRs to examine the local structure of the Ca attachment. Figures 1a,b shows the atomic structures for a Ca atom adsorbed on the edge of a ZGNR with a width of 13.5 Å and an AGNR with a width of 14.4 Å, respectively. The binding energy of the Ca atom on the edge (the middle) of the ZGNR and AGNR are 2.00 (1.10) and 0.73 (0.68) eV, respectively. The calculated binding energy for Ca on the edge of ZGNRs is significantly increased compared to those at elsewhere, and the value for Ca on the different AGNR sites is close to that of ~0.6 eV on graphene or CNTs, which is consistent with the results of the previous study.[36] We find that the transferred charges between the Ca atom and the nanoribbon are more localized when the Ca atom is adsorbed on the edge of a ZGNR than when Ca atom is adsorbed on the middle of the ZGNR as shown in Figure 1c,d. We attribute the larger binding energy of Ca atom on the edge of ZGNRs to the enhanced electrostatic energy by localized transferred charges. Analysis of our results shows that, basically, the binding mechanisms of Ca atom on any hexagonal $sp^2$ carbon structures (e.g., graphene, GNRs, and CNTs) are the same. However, the large binding energy (~2 eV) on the edge of ZGNRs compared to the value of ~0.7−0.6 eV on the middle of ZGNRs, on either the middle or edge of AGNRs, on graphene, and on CNTs is due to a more localized charge transfer on the edge of ZGNRs than in the other cases.

We also consider different geometries of the edge of graphene nanoribbons or large vacancy-defected graphenes with zigzag edges. The binding energy of a Ca atom on an armchair edge for a zigzag-armchair-edged GNR, the zigzag edge of an armchair-zigzag-edged GNR, and the zigzag edge of a large vacancy-defected graphene as shown in Figure 1e–g is 0.90, 1.20, and 1.38 eV, respectively. The binding energy of a Ca atom on the zigzag edges is larger than that on the middle of ZGNRs, on AGNRs, and on graphene. We have also explored the adsorption of Ca atoms on some particular forms of mixed C-BN nanotubes (see Figure 1h,i) because these tubes have the zigzag edges consisting of the carbon atoms. It has been found that the geometries of zigzag C-BN nanotubes depicted in Figure 1h,i are stable.[37] The binding energy of a Ca atom on the zigzag edge (middle) of the carbon region of the (7,0) and (7,7) C-BN nanotubes is calculated to be 1.6 (0.7) and 2.5 (1.6) eV, respectively. Therefore, these results further show that Ca atoms prefer to be adsorbed on the zigzag edge made up of carbon atoms.

We have investigated whether Ca atoms would aggregate on ZGNRs or AGNRs because Ca aggregation changes the bonding nature of adsorption of $H_2$ molecules. Figure 2 shows the optimized atomic geometries for two Ca atoms adsorbed on the edge of a ZGNR (or an AGNR) in the non-aggregated and aggregated structures. Isolated Ca atoms attached to the edge of the ZGNR are preferred in energy over the aggregated case by 0.21 eV (per 2 Ca atoms) because of large binding energy (~2 eV) of Ca atom. In contrast, the aggregation of two Ca atoms on the AGNR is energetically lower than on the non-aggregated case by 0.59 eV (per 2 Ca atoms). These results show that the aggregation of Ca atoms on any hexagonal $sp^2$ carbon structures takes place except for the carbon zigzag edge. To find the lowest energy configuration for two Ca atoms on the edge of a ZGNR, energy minimizing calculations for attachment of an additional Ca atom on the numbered initial positions indicated in Figure 3a are carried out. We find that the configuration of two Ca atoms attached to both sides of one hexagon center of the ZGNR described in Figure 3b is the lowest in energy. Interestingly, the binding energy of the Ca atom is slightly enhanced to 2.2 eV/Ca compared to the value of 2.0 eV/Ca for the case of a single Ca attachment, and this is larger than the cohesive energy of bulk Ca. The increased binding energy of the Ca atoms by 0.2 eV/Ca atom comes from an increased electrostatic interaction by a more localized transferred charge from the Ca atoms to the nanoribbon when they are adsorbed on both sides. To examine how Ca atoms are adsorbed on the edge of ZGNRs, we calculate the total energy for four Ca atoms on the ZGNR as the distance between the Ca atoms that are on the same plane. The relative energy at a Ca-Ca distance of 5.2, 7.7, 10.0, and 12.5 Å is +0.4, +0.2, 0.0, and 0.0 eV, respectively, where the total energy of the lowest energy configuration is set to zero. This result shows that the Ca atoms are repulsive to each other, and the minimum Ca-Ca distance with negligible interaction between them is 10 Å as shown in Figure 3c. Therefore, a maximum Ca coverage on ZGNRs can be obtained with a Ca-Ca distance of 10 Å without a clustering of Ca atoms.

Next, we consider boron doping on AGNRs to suppress Ca aggregation by preferential Ca binding to B sites. We find that the B atoms prefer to be on the edge of AGNRs as distant from each other as possible. The minimum B-B distance without significant interaction between the B atoms is ~9 Å. The binding energy of the Ca atom on the hexagon center including the B atom is increased considerably to 2.2 eV compared to ~0.7 eV on pristine AGNRs. As in the case of B-doped CNTs, with boron decoration, the non-aggregated case of Ca atoms is more favorable in energy by 0.13 eV (per 2 Ca atoms) than the aggregation case as shown in Figure 3d,e. These results show that Ca atoms prefer to be individually adsorbed on the B sites of B-doped AGNRs without an aggregation of Ca atoms. Unlike AGNRs, the binding energy of Ca atom on B-doped ZGNRs is slightly reduced to ~1.8 eV/Ca compared to the value ~2.0 eV/Ca for pristine ZGNRs. However, Ca clustering is still suppressed on B-doped ZGNRs.

We calculate the binding energy of $H_2$ molecules on the Ca atom as a function of the number of adsorbed $H_2$ molecules for an isolated Ca atom adsorbed on a ZGNR (B-doped AGNR). Figure 4a,b shows the optimized atomic structures for the configuration with the number of maximally adsorbed $H_2$ molecules on the Ca atom attached to the ZGNR and B-doped AGNR, respectively. The calculated (average) binding energies of the $H_2$ molecules on the Ca atom attached to the ZGNR (B-doped AGNR) with the GGA are 0.18 (0.20), 0.21 (0.20), 0.20 (0.18), 0.17 (0.18), 0.16 (0.15), and 0.13 (0.15) eV/$H_2$ for 1, 2, 3, 4, 5, and 6 $H_2$ molecules, respectively. The calculated binding energies with the local density approximation (LDA) are approximately twice as large as those with the GGA. The distance between the Ca atom and the $H_2$ molecule is ~2.5 Å and the bond length of the $H_2$ molecule is slightly elongated to ~0.77 Å from 0.75 Å of the isolated molecule. According to a recent study,[36] the configuration with opposite spin orientation between ferromagnetically ordered edges is more stable than the one with the same spin orientation between the two edges regardless of alkali or alkali earth metal adsorption. The further adsorption of $H_2$ molecules also does not affect the magnetic ground state because the binding mechanism of the $H_2$ molecules on the Ca comes from the polarization of the $H_2$ molecules. We believe that the magnetic structure is not expected to influence the conclusion of the paper.

Figure 4c,d shows the optimized atomic structures for a maximum hydrogen storage capacity in a Ca-decorated ZGNR and B-doped AGNR with the Ca-Ca minimum distance of 10 and 9 Å within the same edge that corresponds to negligible interaction between the Ca atoms, respectively. Since the binding energy of Ca atoms (~2 eV/Ca) on the edges of ZGNRs is much larger than the value (~0.7 eV) elsewhere, Ca atoms are favorably adsorbed on the edges. So, only adsorption of Ca atoms on the edges of the nanoribbons is considered. Furthermore, when Ca atoms are attached to both sides of ZGNRs, the maximum number of adsorbed $H_2$ molecules is reduced to 5 $H_2$ molecules compared to a maximum of 6 $H_2$ when a single Ca atom is attached to one side. The hydrogen storage gravimetric capacity reaches 5.1 and 6.5 wt % hydrogen, respectively. The molecular formula for the Ca-decorated ZGNR and B-doped AGNR may be expressed as $(C_{48}H_8 \cdot 4Ca \cdot 20H_2)_n$ and $(C_{42}B_2H_8 \cdot 4Ca \cdot 24H_2)_n$ (n is an integer), respectively. A recent paper on a Ca-decorated graphene reported the maximum hydrogen storage capacity as high as 8 wt%.[31]

Our calculations suggest the potential of Ca-decorated graphene-based nanostructures such as GNRs as a hydrogen storage medium. In recent years, graphene-based nanostructures have received a lot of attention and much effort has been devoted to making various structures because of their interesting physical properties.[38,39] Very recently, graphene nanoribbons have been fabricated experimentally from CNTs.[40,41] Furthermore, wide grapheme samples (~2cm×2cm) have been synthesized using catalysts[42] and large vacancy-defected graphenes have been made[43], and C-BN nanotubes have been synthesized as well.[44] These experiments indicate that Ca-decorated zigzag graphene-based nanostructures we propose here can be made.

In conclusion, we have demonstrated the possibility that individually dispersed Ca-decorated graphene-based nanostructures can serve as a high-capacity hydrogen storage medium. Ca clustering is suppressed on the zigzag edge and on B-doped armchair edge of graphene. We feel that these systems can be made, and we encourage experimental searches to synthesize these hydrogen storage nanomaterials that may operate at room temperature and ambient pressure.

**ACKNOWLEDGMENT.** This research was supported by the National Science Foundation Grant No. DMR07-05941 and by the Director, Office of Science, Office of Basic Energy Sciences, Materials Sciences and Engineering Division, U. S. Department of Energy under Contract No. DE-AC02-05CH11231. Computational resources were provided by NPACI and NERSC. J. Ihm was supported by the Center for Nanotubes and Nanostructured Composites funded by the Korean Government MOST/KOSEF, and the Korean Government MOEHRD, Basic Research Fund No. KRF-2006-341-C000015.

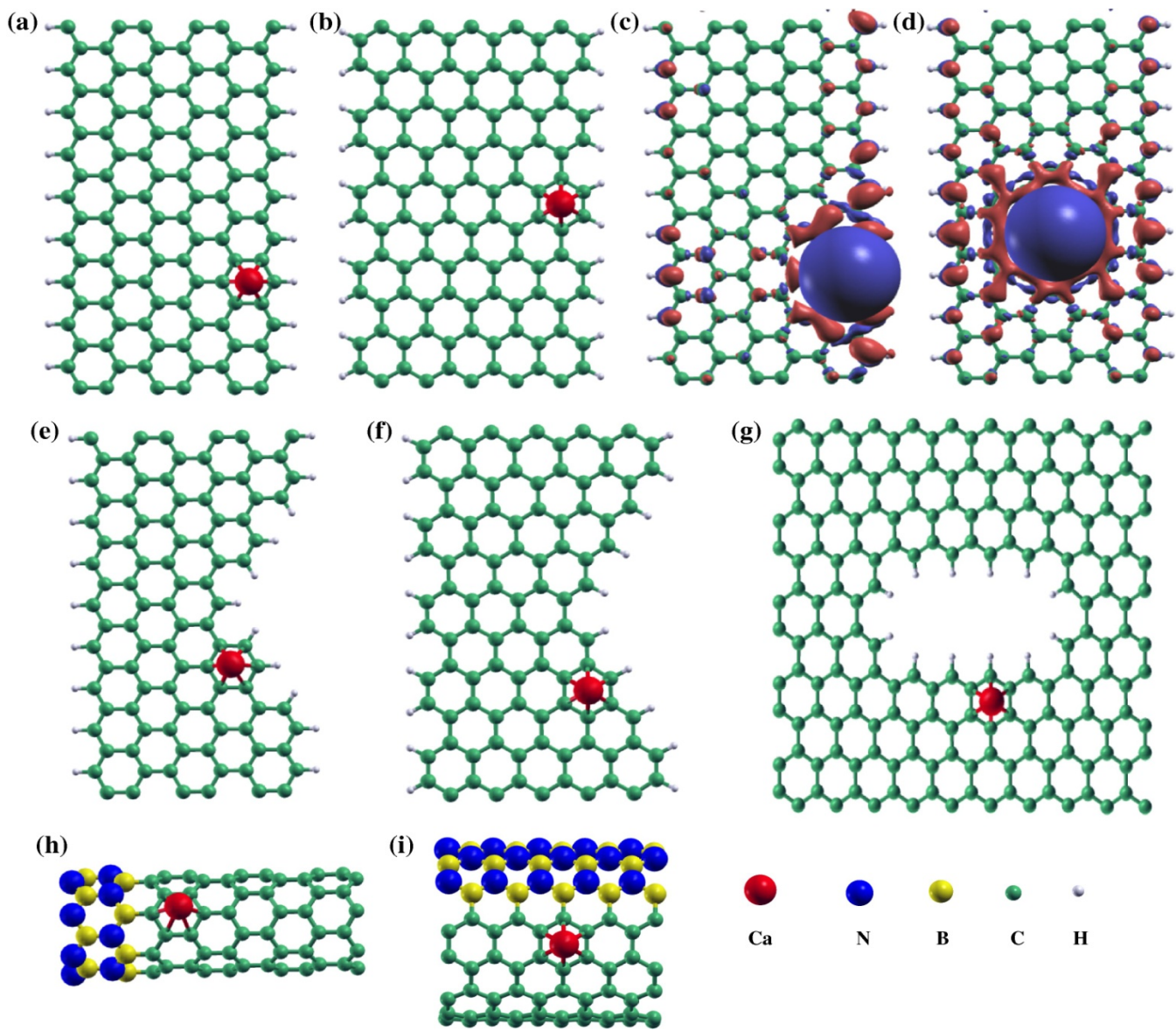

**Figure 1.** (a) and (b) show the optimized atomic geometries for a Ca atom adsorbed on the edge of a ZGNR and AGNR, respectively. (c) and (d) show the charge density difference between the Ca atom and the ZGNR with the isosurface value of 0.0005 $e/(a.u.)^3$ when the Ca atom is attached on the edge and the middle, respectively. Red and blue colors indicate electron accumulation and depletion, respectively. (e)–(i) show the optimized atomic structures of a Ca atom adsorbed on the armchair edge of a zigzag-armchair-edged GNR, the zigzag edge of an armchair-zigzag-edged GNR, the zigzag edge of a large vacancy-defected graphene, a (7,0) C-BN nanotube junction, and a (7,7) mixed C-BN nanotube, respectively.

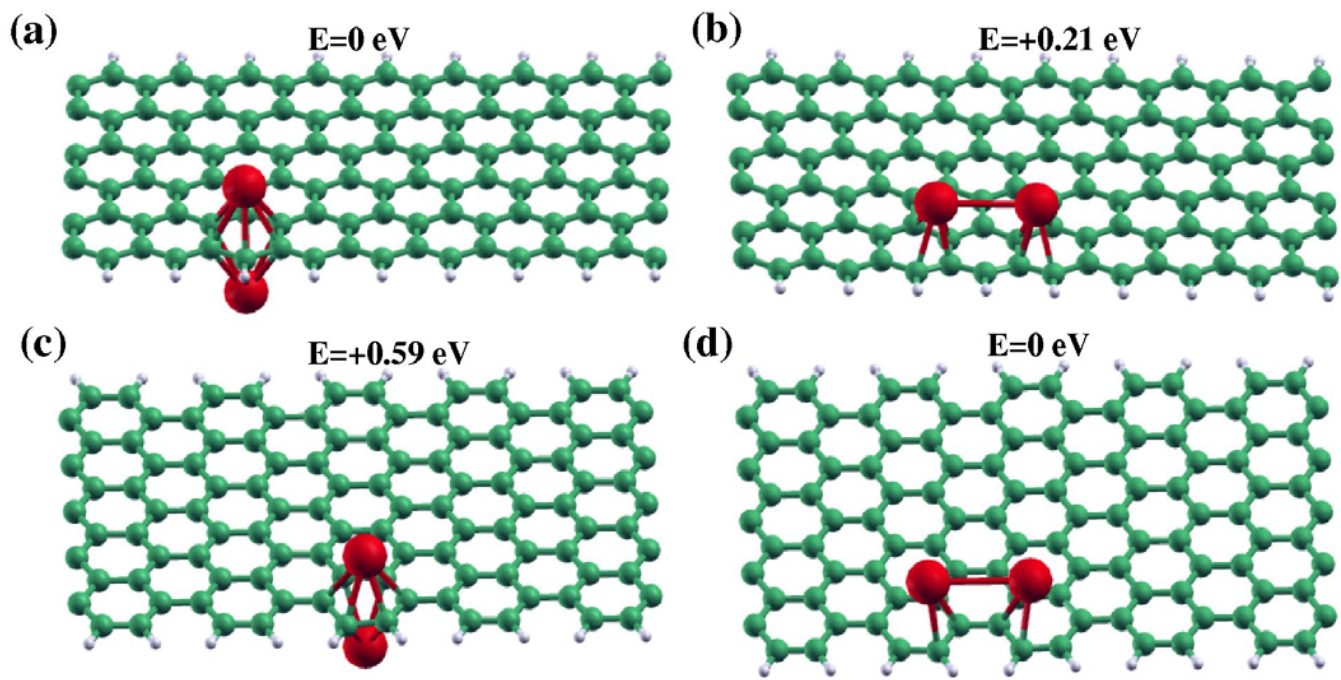

**Figure 2.** (a) Two Ca atoms individually attached on the both sides of the edge of a ZGNR. (b) Two Ca atoms aggregated on the edge of a ZGNR. (c) Two Ca atoms individually attached on the edge of an AGNR. (d) Two Ca atoms aggregated on the edge of an AGNR. The energy (E) means the total energy where the lower energy between aggregated and non-aggregated structure is set to zero.

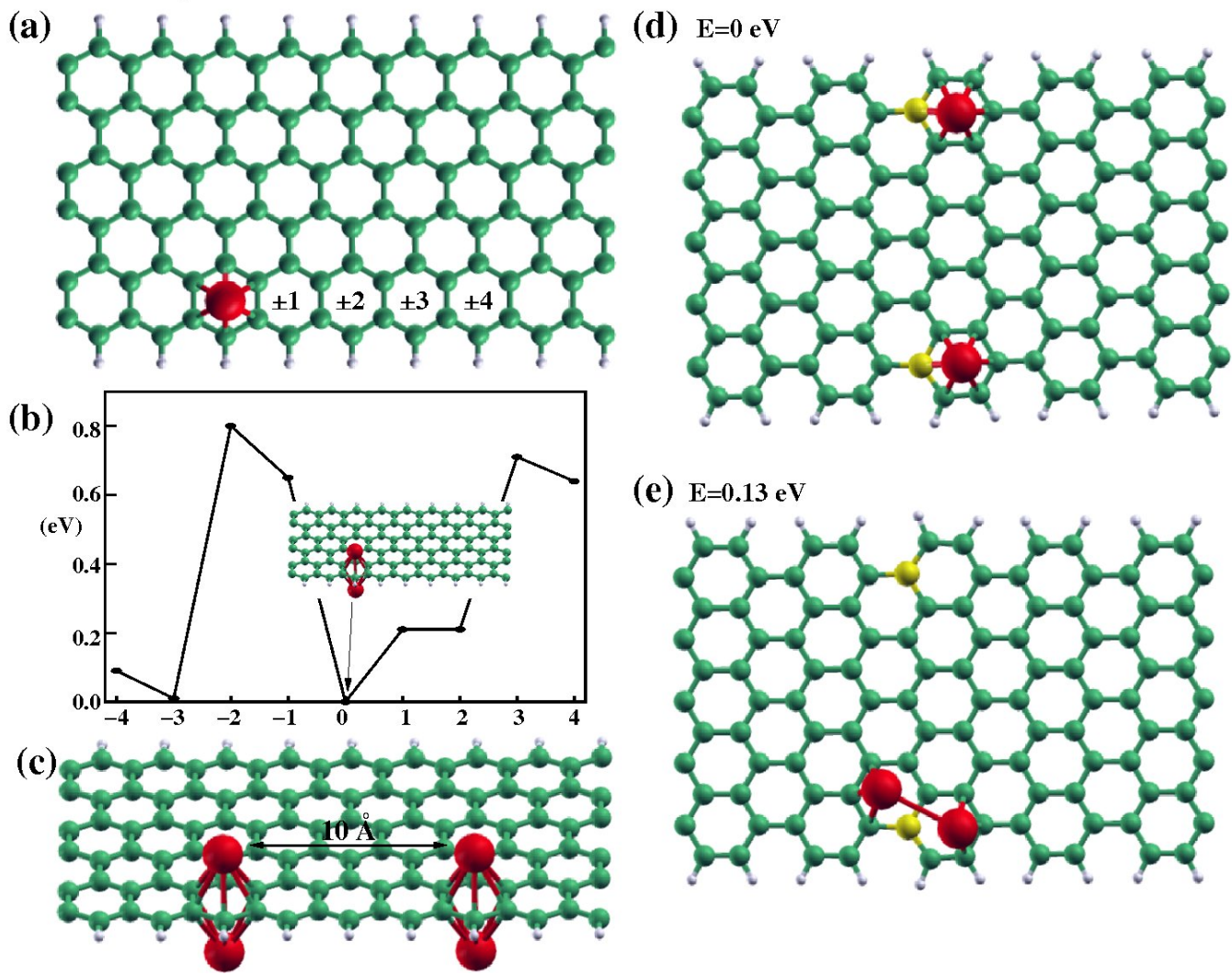

**Figure 3.** (a) Initial positions for the attachment of an additional Ca atom on the edge with adsorbed Ca atom of a ZGNR. + (–) sign means the same (opposite) plane with the Ca atom attached. (b) The total energy of the optimized structures for two Ca atoms on the edge of the ZGNR as a function of different initial positions presented in Fig. 3(a) where the energy of the lowest energy configuration is set to zero. "0" means the opposite site of the attached Ca atom. Inset shows the lowest energy configuration. (c) Optimized atomic structure of the lowest energy configuration for the attachment of four Ca atoms with a distance of 10 Å between the Ca atoms that are on the same plane. (d) Two Ca atoms individually attached on the edge of a B-doped AGNR. (e) Two Ca atoms aggregated on the edge of a B-doped AGNR. The total energy (E) of the lower energy structure between aggregated and non-aggregated structures is set to zero.

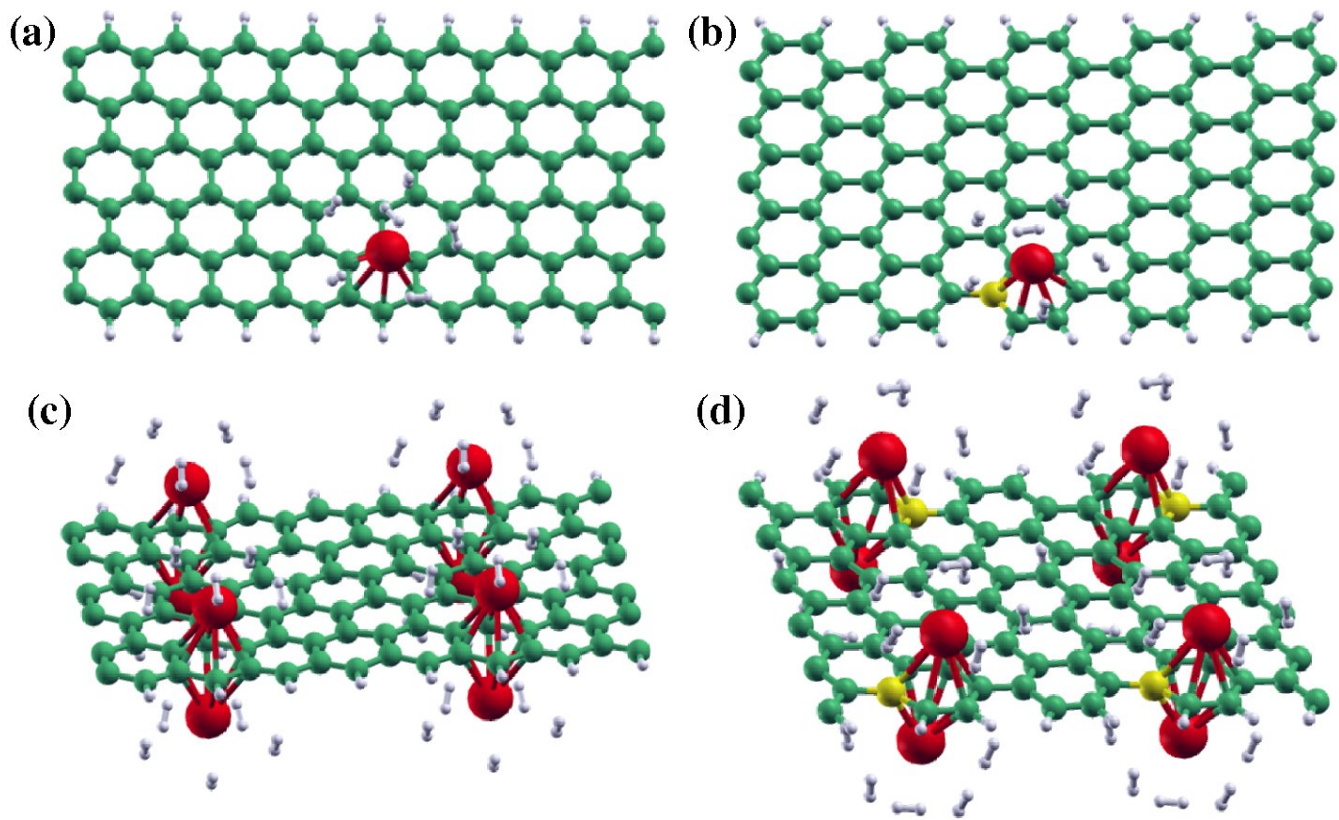

**Figure 4.** (a) and (b) show the optimized atomic geometries of a Ca atom adsorbed on the edge of a ZGNR and B-doped AGNR with maximum number of $H_2$ molecules, respectively. (c) and (d) show the optimized atomic geometries of maximum number of adsorbed $H_2$ molecules for a Ca-decorated ZGNR and B-doped AGNR (4.5 at. % of B), respectively.